# Design, Characterization and Indoor Validation of the Optical Soiling Detector "DUSST"

Alvaro F. Solas[a, *], Leonardo Micheli[a], Matthew Muller[b], Florencia Almonacid[a], Eduardo F.Fernandez[a]

[a] Photovoltaic Technology Research Group (PVTech-UJA), CEACTEMA, University of Jaén (UJA), Las Lagunillas Campus, Jaén, 23071, Spain

[b] National Renewable Energy Laboratory (NREL), 15013 Denver West Parkway, Golden, Colorado 80401 USA

## Abstract

Nowadays, photovoltaic (PV) technology has reached a high level of maturity in terms of module efficiency and cost competitiveness in comparison with other energy technologies. As PV has achieved high levels of deployment, the development of devices that can help to reduce PV operation and maintenance costs has become a priority. Soiling can be cause of significant losses in certain PV plants and its detection has become essential to ensure a correct mitigation. For this reason, accurate and low-cost monitoring devices are needed. While soiling stations have been traditionally employed to measure the impact of soiling, their high cost and maintenance have led to the development of innovative low-cost optical sensors, such as the device presented in this work and named "DUSST" (Detector Unit for Soiling Spectral Transmittance)". The thermal characterization of DUSST's components and the methodology used to predict soiling transmittance losses are presented in this study. The results show that the losses can be predicted with an error lower than 1.4%. The method has been verified with an experimental campaign with naturally soiled coupons exposed outdoors in Jaén, Spain.

## Nomenclature

*Symbols*

| | |
|---|---|
| $I_{SC}$ | Short-circuit current [A] |
| $P_m$ | Maximum power [W] |
| SR | Soiling Ratio |
| $T(\lambda)$ | Hemispherical transmittance at wavelength $\lambda$ |

*Greek letters*

| | |
|---|---|
| $\lambda$ | Wavelength [nm] |

*Abbreviations*

| a-Si | Amorphous Silicon |
|------|-------------------|
| AST | Average Transmittance of Soiling |
| CdTe | Cadmium Telluride |
| IEC | International Electrotechnical Commission |
| LIR | Light Intensity Ratio/ % |
| MAE | Mean Absolute Error/ % |
| m-Si | Monocrystalline Silicon |
| PM | Particulate Matter |
| PMMA | Polymethyl methacrylate |
| PVGIS | Photovoltaic Geographical Information System |
| $R^2$ | Coefficient of Determination |
| RMSE | Root Mean Square Error |
| UV-VIS | Ultraviolet-Visible |
| TLED_Factor | Factor of LED Temperature |

*Subscripts*

| baseline | reference (clean) conditions |
|----------|------------------------------|
| c | cell |
| i | a counting index |
| loss | losses due to soiling |
| measured | measured data point |
| modelled | modelled data point |

## 1. Introduction

Soiling is a natural phenomenon that impacts photovoltaics (PV) systems worldwide. It consists of the deposition of dust, dirt and other contaminant particles on PV modules. Soiling reduces the intensity of the solar radiation that can reach the PV cells which directly translates into a reduction of the output power of solar modules. Several reviews have addressed the impact of soiling on PV systems (Costa et al., 2018, 2016; Sayyah et al., 2014). In order to mitigate and reduce these effects, which may lead to significant energy losses, and thus result in high income reductions (Ilse et al., 2019), monitoring soiling losses has become important for the PV community. As a consequence, there has been remarkable growth in both the deployment of devices that monitor and quantify soiling losses as well as the development of different methods to extract them directly from PV performance (Deceglie et al., 2018; Figgis et al., 2017; Kimber et al., 2006; Micheli et al., 2020a). Deceglie et al. (2018) presented a method that allows the automatic estimation of soiling losses from PV data and validated it against the results obtained with traditional measurement systems. Figgis et al. (2017) reviewed the different approaches to measure and evaluate the impact of soiling in PV systems. Micheli et al. (2020a) introduced a method to extract the soiling profile and quantify the seasonal variation in soiling using PV and precipitation data.

The most common soiling monitoring solutions are the so-called soiling stations. These consist of two PV devices: one of them must be regularly cleaned and the other is left to naturally soil. Soiling losses are calculated by the comparison of the electrical output of both devices through the soiling ratio (SR) index, as stated in the IEC 61724-1 Standard (International Electrotechnical Commission, 2017). It is expressed as:

$$SR = \frac{Z_{soiled}}{Z_{baseline}}, \qquad\qquad (1)$$

where $Z_{soiled}$ and $Z_{baseline}$ are the electrical outputs of the soiled and clean devices, respectively. The electrical output can be either the short-circuit current ($I_{sc}$) or the maximum power ($P_m$). SR has a value of 1 in clean conditions, whereas it lowers with the accumulation of soiling. In other words, the higher the soiling accumulation (and the soiling loss), the lower the soiling ratio.

Despite their simple measurement procedure, soiling stations present some disadvantages. They require frequent maintenance and even slight misalignments between the two devices may cause high uncertainties in the results (Gostein et al., 2014; Micheli et al., 2017; Muller et al., 2017). To overcome these drawbacks, innovative, low-cost and low-maintenance optical soiling sensors are being developed. Figgis et al. (2016) converted a digital microscope into a soiling sensor able to quantify soiling and to detect also condensation. Valerino et al. (2020) used a similar setup to estimate soiling losses using empirical models. Commercial optical soiling sensors have also become available in the market. The sensor presented by Gostein et al. (2018), named Mars and developed by Atonometrics, uses a camera to take pictures of a soiled glass and an image analysis process to determine the transmittance of the dust layer. On the other hand, the device introduced by Korevaar et al. (2017), called DustIQ and developed by Kipp & Zonen, estimates soiling losses through changes in reflectance measurements. In this case, a photodiode measures the amount of light emitted from a LED and reflected from the dust particles accumulated on the surface of a glass coupon. A brief comparison of different solutions for soiling monitoring in PV systems is given in Table 1.

Recently, an innovative optical soiling detector named DUSST (Detector Unit for Soiling Spectral Transmittance) was designed to monitor soiling without moving parts, the action of an operator and the need of water. It has been developed and patented by researchers of the University of Jaén (Spain) and National Renewable Energy Laboratory, NREL (CO, USA) (Fernández et al., 2019). DUSST quantifies the soiling losses by comparing the light transmitted through a soiled glass with the light transmitted through the same glass in clean conditions.

*Table 1. Comparison of soiling monitoring solutions.*

| Name | Description | Reference |
|------|-------------|-----------|
| **DUSST** | It is an optical sensor that uses a monochromatic light source and a PV cell as light detector. Soiling losses are quantified by comparing the light transmitted through a soiled glass with the light transmitted through the same glass in clean conditions. | Fernández et al., 2019 |
| **Traditional soiling station** | It consists of two identical PV devices: one of them must be regularly cleaned and the other is left to naturally soil. Soiling losses are calculated by the comparison of the electrical output of both devices. | Gostein et al., 2015 |
| **Waterless soiling station** | It follows the same operation principle as a traditional soiling station. Regular cleanings are avoided by the use of a glass shutter that covers the PV device that should remain clean. | Curtis et al., 2018 |
| **Mars** | It is an optical soiling sensor that uses a camera to take images of a soiled glass. Soiling transmittance losses are estimated by image analysis techniques. | Gostein et al., 2018 |
| **DustIQ** | It is an optical sensor that uses a photodiode, which measures the amount of light emitted from a LED and reflected from the dust particles accumulated on the surface of a glass coupon. A specific calibration of the sensor should be conducted for each location. | Korevaar et al., 2017 |
| **Outdoor soiling microscope** | Soiling losses are calculated by obtaining the fraction of area of a glass coupon that is covered with dust (surface coverage). Soiling rate is obtained by the difference in surface coverage between successive images. | Figgis et al., 2016 |
| **Low cost digital microscope** | Similar setup as the one presented by Figgis et al. (2016). Soiling losses are estimated by using an optical model. Results were validated against a traditional soiling station. | Valerino et al., 2020 |

In this work, the results presented by Muller et al. (2018) are extended, by adding a thermal characterization performed under controlled conditions and by presenting an empirical model to estimate the soiling transmittance losses with DUSST. These findings will be used to improve the robustness and reliability of the outdoor measurements taken by DUSST, facilitating its field deployment. In addition, a complete description of the components, operation principles and fundamentals of DUSST is provided.

The paper is structured as follows: Section 2 presents the components of DUSST and describes its principles and daily operation; Section 3 details the materials and methods that have been employed to perform the experiments; Section 4 shows the methodology and the results of an indoor thermal characterization of DUSST; Section 5 presents and analyses the model that allows the estimation of transmittance losses due to soiling; finally, Section 6 recapitulates the main conclusions found during the development of this study.

## 2. Sensor description

### 2.1. DUSST principles and components

DUSST measures the collimated monochromatic light emitted from a stable source and transmitted through a naturally soiled glass to quantify the soiling ratio of a PV array. Fig. 1 shows the block diagram of DUSST and the version of the indoor prototype that was used during the experiments. This prototype incorporates some improvements in comparison with the one presented by Muller et al. (2018), such as: a better collimation of light onto the detector and an increase of its robustness through the use of weather-resistant materials for outdoor deployment.

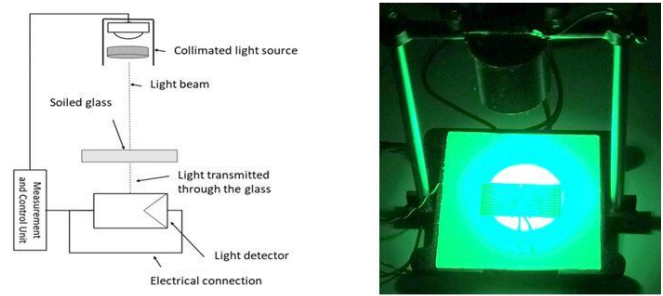

*Fig. 1. Left: block diagram of DUSST prototype. Right: indoor DUSST prototype.*

Its main components are: (1) the collimated light source that consists of a monochromatic light-emitting diode (LED) emitting light at 530 nm (green) and an optical element to collimate the light onto the light detector; (2) an encapsulated PV cell as light detector (cell + PV glass).

The choice of a green monochromatic light source (emitting at 530 nm) is based on the study presented by Micheli et al. (2019). In that experimental work, it was proven that the impact of soiling on the electrical output of PV cells can be estimated, with high accuracy, by using single-wavelength optical transmittance measurements between 500 nm and 600 nm. The soiling transmittance was calculated by comparing the spectral transmittance of a glass coupon mounted in outdoor conditions and the spectral transmittance of a clean glass, as follows:

$$T_{soiling}\ (\lambda) = \frac{T_{soil}\ (\lambda)}{T_{baseline}\ (\lambda)}, \tag{2}$$

where $T_{soil}\ (\lambda)$ and $T_{baseline}\ (\lambda)$ are the hemispherical transmittance of the soiled coupon and reference coupon, respectively, for one specific wavelength $(\lambda)$.

The soiling ratio values for different PV technologies were estimated by combining the spectral transmittance of soiling, the global spectral irradiance and the spectral response of each PV technology. The authors found clear linear correlations between the soiling ratios and the average transmittance of soiling (AST($\lambda$)). The AST($\lambda$) represents the mean soiling transmittance over a specific spectral waveband, and was estimated for different regions of the spectrum (Ultraviolet, Visible and Near-Infrared) and for the spectral response of different PV technologies. The best linear fits for the three different PV technologies represented in Fig. 2 were found for the visible region of the spectrum, which covers wavelengths between 400 nm and 700 nm. In this range, the coefficient of determination ($R^2$) values were higher than 0.93 in all cases; whereas in the near-infrared region [700 nm – 1240 nm] and especially in the ultraviolet region [300 nm – 400 nm], $R^2$ values were lower.

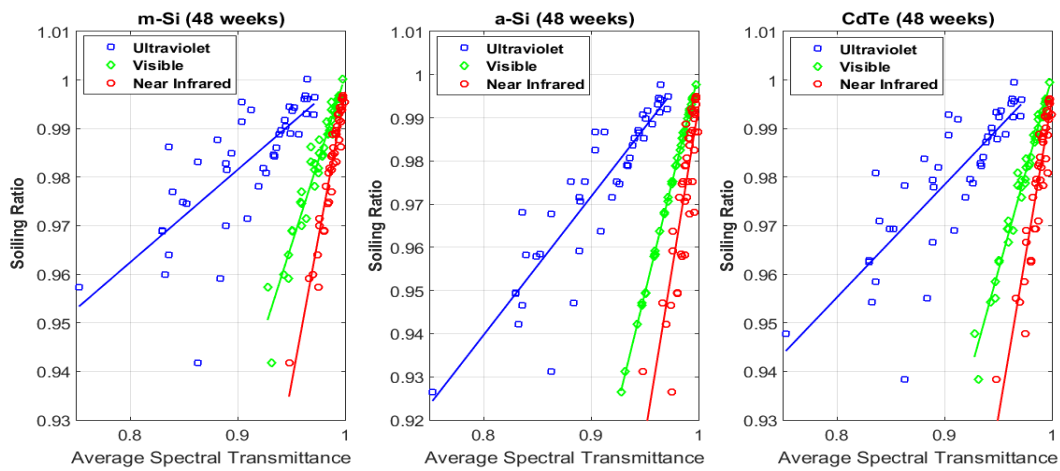

*Fig. 2. Average Spectral Transmittance for three regions of the spectrum vs. Soiling Ratio modelled from the hemispherical transmittance, the spectral irradiance and the spectral response of three PV technologies. Adapted from* (Micheli et al., 2019).

In addition, the authors of (Micheli et al., 2019) investigated the possibility of estimating the soiling ratio using single-wavelength transmittance measurements. This was performed by considering wavelengths between 300 nm and 1000 nm. It was found that the best estimations ($R^2 > 0.9$) were obtained for transmittances measured at wavelengths between 500 nm and 600 nm for various PV technologies, see Fig. 3. For this reason, a green-coloured diode, emitting light at 530 nm, was selected for the prototype of DUSST presented in this work.

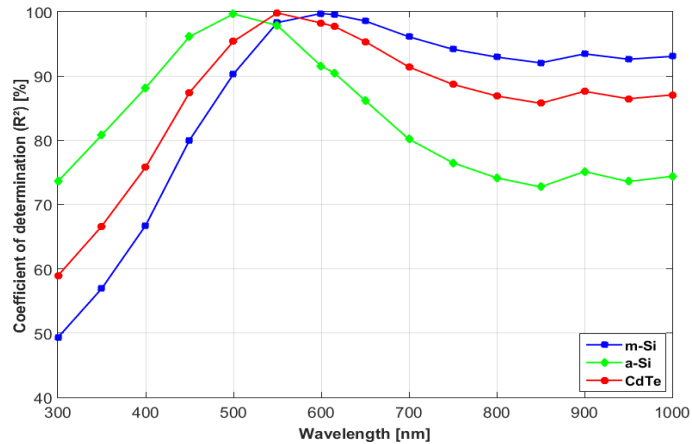

*Fig. 3. $R^2$ obtained when soiling losses for various PV technologies located in Jaén, Spain are estimated using a single hemispherical transmittance wavelength.*

All the currently available soiling sensors, including the soiling stations report the losses for a single PV technology. While this approach provides reasonable results, as the soiling trends will be similar for the various PV technologies, DUSST is readily adaptable to the different semi-conductor devices. As demonstrated in (Micheli et al., 2019), the colour of the LED can be changed depending on the PV technology, where high-energy-band semi-conductors can be better represented by shorter wavelength LEDs and low-energy-band semi-conductors by higher wavelength LEDs. In addition, it should be noted that DUSST can be built with multiple LEDs of separate wavelengths in order to model the full spectral transmittance curve, achieving even a better soiling detection, as discussed in (Micheli et al., 2020b). This is being investigated and will be addressed in future works. This paper only focuses on a single-measurement approach and therefore describes the main features and the basic operation principles of DUSST with a 530 nm LED.

It is important to highlight that the light detector of DUSST mimics a typical PV module. Indeed, in DUSST, the glass that is allowed to soil is light-coupled to a PV cell with ethylene vinyl acetate, as is done with full size PV modules. This makes DUSST different from other soiling sensors that only use a stand-alone glass coupon, with soiling mechanisms potentially different from those of a PV module. This way, the soiling losses provided by DUSST are expected to be similar to fielded PV modules.

## 2.2. DUSST operation

In this subsection, the daily DUSST measurement procedure is described. DUSST quantifies the daily soiling losses by analyzing the soiling ratios measured at multiple moments each night, when the influence of external light is negligible. Several measurements are taken to reduce the impact of variable external light conditions in the surroundings, which can bias the reading. Each measurement consists of five steps:

1. **Zero Measurement**: the intensity of the light detector is recorded while the collimated light source is turned off. Thus, the influence of external light sources, which might affect the soiling measurement, is quantified. Minimal external light noise is corrected, whereas, in conditions of extreme external light, the measurements are not taken.

2. **Soiling Measurement**: the collimated light source is switched on. Once the signal is stable, the intensity of the light measured by the light detector is recorded.

3. **Soiling Measurement Correction**: the soiling measurement is corrected by subtracting the zero-measurement to remove the effect of external light sources.

4. **Light Intensity Ratio (LIR)**: the ratio between the intensity of the light transmitted through the soiled glass and its baseline (intensity of the light transmitted through the clean glass) is calculated as:

$$\text{LIR [\%]} = \frac{I}{I_{\text{baseline}}} \times 100\%, \tag{3}$$

where $I$ is the output current in soiled conditions and $I_{\text{baseline}}$ is the output current in clean conditions. The more soil accumulated, the lesser its value since $I$ diminishes with soiling. Subsequently, the electrical losses measured with DUSST can be calculated through the following equation:

$$\text{DUSST}_{\text{losses}} \text{ [\%]} = 100\% - \text{LIR}. \tag{4}$$

The Light Intensity Ratio and the electrical losses measured by DUSST can be used to estimate the transmittance losses due to soiling. This correlation is one of the focuses of this study.

5. **Soiling Ratio**: the soiling ratio is calculated by correcting the value obtained in step 4 according to theoretical coefficients that take into account the type of PV modules under investigation as described in (Micheli et al., 2019). It can be expressed as:

$$\text{SR}_i [\%] = \text{Slope}_i \times \text{DUSST}_{\text{losses}} \text{ [\%]} + \text{Offset}_i, \tag{5}$$

where Slope and Offset are calibration coefficients that relate the output of DUSST and the soiling losses of a certain *i*-PV technology, which can been estimated by conducting analyses such as those shown in Fig. 2.

The baseline measurement, which represents the intensity of the light transmitted through the glass in a clean condition, is determined (1) at the moment the device is installed, and (2) after each cleaning. Indeed, the detector needs to be cleaned in parallel with any cleanings performed for the PV systems it is monitoring. Thanks to this periodic update of the baseline measurement, possible influences of other types of degradation on the soiling measurement, such as aging of DUSST components, are limited.

### 3. <u>Materials and Methods</u>

3.1. Monitoring system

The experiments described in this study have been performed at the CEACTEMA solar energy laboratory of the University of Jaén, in Jaén, Spain (latitude 37º49'N, longitude 3º48'W), under controlled conditions of temperature (24 ºC ± 1 ºC), relative humidity (40 % ± 5 %) and minimal external light intensity. In order to characterize the thermal behaviour of the different components of DUSST, their temperature and electrical parameters while in operation have been monitored. A block diagram for the system hardware is shown in Fig. 4. The equipment, which has been used, is listed below:

1) An *Agilent E3631A DC* power supply. It was set to a fixed voltage of 12 V to power the light source. The current was limited to 1 A.
2) A monochromatic green (530 nm) power LED as the light source. It nominally produces 158 lumens of light for an operating current of 700 mA.
3) An optical element to collimate the light beam from the LED with an angle of 9º.
4) A LED power module (*LUXEON Buckpuck 3021*), to ensure that a constant current has been applied to the light source.
5) A poly-crystalline silicon PV solar cell as light sensor.
6) An *Agilent 34970A* data acquisition (DAQ) unit and a *34901A channel multiplexer* to register the electrical parameters of the different components and temperatures. This multiplexer has 20 channels of general purpose and 2 additional fused inputs that allow current measurements up to 1 A without the need for external shunt resistors.
7) 3 type J thermocouples mounted on the back of the LED, on the encapsulation of the LED power module and on the rear side of the PV solar cell. This type of thermocouple can measure temperatures up to 260 ºC with a time response over 1 s and according to IEC-584, the tolerance is class 1.

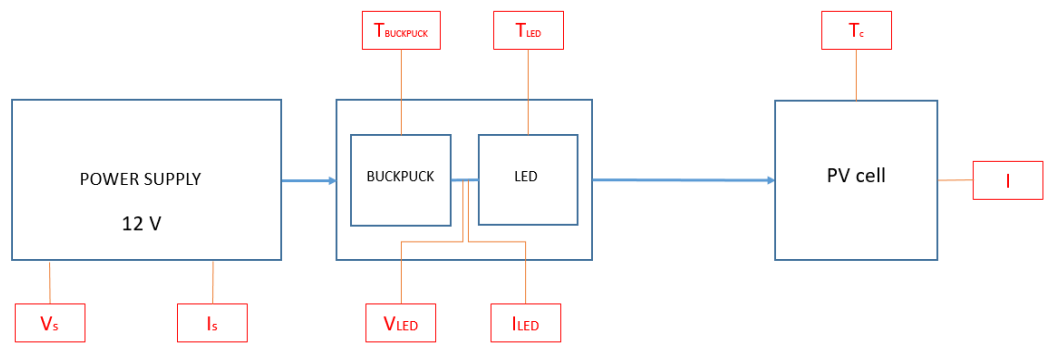

*Fig. 4. Block diagram of the monitoring system.*

LabVIEW software has been used to develop a control and acquisition program to register and save all the data measured by the different sensors.

### 3.2. Soiling data collection

Two sets of data were collected to conduct the studies presented in this work. First, during the calibration stage, a set of 12 screening masks with different levels of grey were printed with a high-quality laser printer on a transparent polyester film. Each mask had a size of 60 mm x 80 mm. 12 different shades of grey, ranging from 0% to 100% black percentage, at steps of 5%, were used to simulate different levels of soiling density. It should be mentioned that the black percentage does not directly correspond to the transmittance of the mask, as this, indeed, depends also on other additional factors, such as the material's own transmittance. The models presented in section 5 were obtained by using these printed masks. This makes the models independent of the chemical and physical properties of dust, thus allowing for their application in any location.

After this first stage, 12 PMMA PLEXIGLAS$^{®}$ 8N clear coupons 60 mm x 80 mm in size and 6 mm thick were mounted outdoors on the roof of the C6-building at the University of Jaén. The coupons were mounted in the summer dry season when moderate soiling was expected. Jaén is a city located in the southeast of Spain, with a high annual solar irradiation (1778 kWh/m² on horizontal plane according to PVGIS-5 geo-temporal irradiation database (European Commision, 2016)) and a Mediterranean-continental climate (Nofuentes et al., 2017). Annual average values of airborne particulate matter are usually low-medium. However, due to its proximity to the Sahara Desert, high values of dust density can be reached when two conditions are met: (1) wind speed is moderately high, lifting the sand from the dunes and (2) wind direction is 180º (wind blows from the south). Additionally, the coupons were placed horizontal in order to maximize the accumulation of soiling on their surface. They were removed from the roof at different times, with variable intervals between 4 and 14 days, depending on the meteorological conditions and airborne particulate matter ($PM_{10}$) levels.

The use of coupons to analyze soiling losses is widely spread among the PV community. In fact, most of the innovative optical soiling detectors mentioned before use glass coupons during their operation. In this study, coupons are only used for the validation of the model that correlates the electrical measurements of DUSST and the soiling transmittance losses.

### 3.3. Methodology

In the work presented by Muller et al. (2018), a preliminary model that correlates soiling transmittance losses with the measurements of the first DUSST prototype was presented. This present study aims to obtain and validate a novel model to accurately estimate outdoor soiling losses independently of the temperature conditions. In this subsection, the methodology that was followed to develop the model is presented.

The experimental process was divided in two different stages: the first stage consisted of the measurement and calibration of the PV cell response using the printed masks, whereas the second stage involved the validation of the mask-based per the coupons that were allowed to soil naturally outdoors. In both cases, the data acquisition process was the same. This process is summarized in Fig. 5.

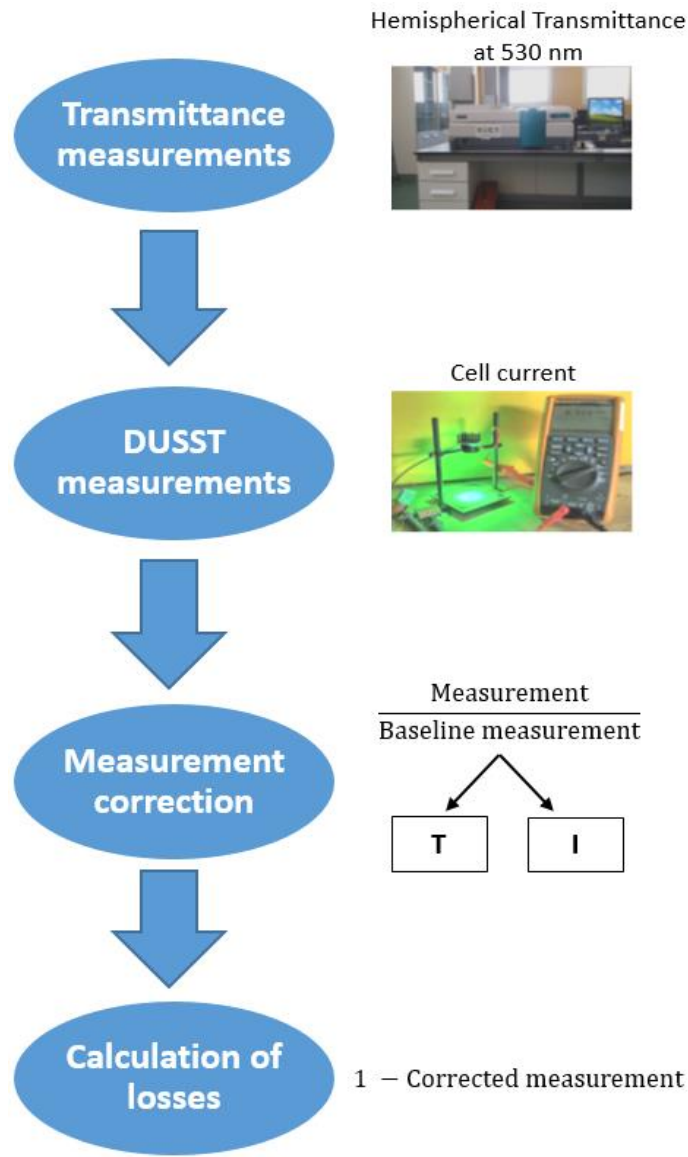

*Fig. 5. Data acquisition process.*

The hemispherical transmittances of the samples were measured at 530 nm using a CaryWin 4000 dual-beam ultraviolet-visible (UV-VIS) spectrophotometer located at the Centre of Scientific-Technological Instrumentation of the University of Jaén. The transmittance of a clean coupon (reference coupon), which was measured at the beginning and at the end of the experiment, was used as a baseline and was employed in equation 6 to calculate the value of the soiling transmittance. This reference coupon was stored in a dry and clean environment to prevent the accumulation of soiling on its surface. There were some differences between transmittance measurements conducted at the calibration stage and the ones conducted during the validation stage. First, during the calibration stage, the screening masks were fixed to the surface of the reference coupon; whereas during the validation stage only the transmittance of the naturally soiled coupons was measured. Second, to take into account the presence of non-uniform soiling over naturally soiled coupons, four transmittance measurements were performed in different points of their surface during the validation stage. The average of these measurements was calculated. Also the standard deviation was calculated as a way to quantify the level of non-uniformity.

In this work, the DUSST measurements was conducted by recording the PV cell current and LED temperature during a 21-minute period with a 2-seconds sampling interval. Then, the data were processed to obtain a corrected value of current, independently of the LED temperature and external conditions. This methodology is shown in Fig. 6. It should be noted that, in order to minimize the impact of outliers, when the difference between two consecutive cell current measurements was higher than 0.2 mA, the measurements was replaced with the average of the 5 previous and 5 posterior measurements.

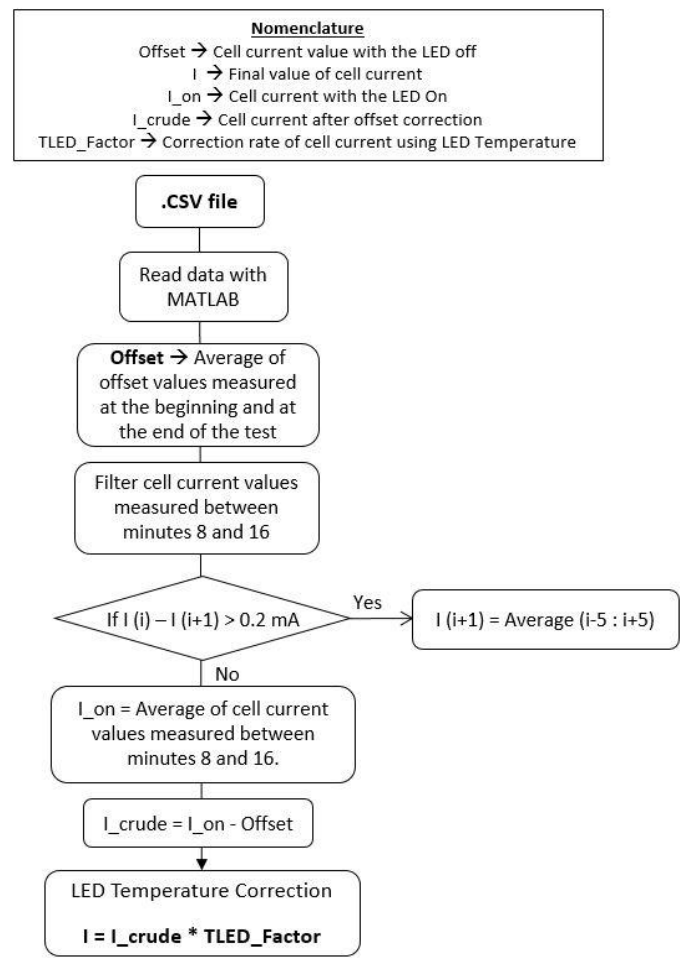

*Fig. 6. Methodology to process the DUSST measurements.*

Once the transmittance measurements and DUSST measurements were taken, they were normalized to the reference coupon measurement (baseline). DUSST measurements were standardized using equation 3, whereas equation 6 was used for transmittance measurements, similar to equation 2:

$$T_{soiling} \ (530 \ nm) \ [\%] = \frac{T_{measured \ (530 \ nm)}}{T_{baseline \ (530 \ nm)}} \times 100\%, \tag{6}$$

where $T_{soiling}$ (530 nm) is the soiling transmittance at 530 nm; and $T_{measured}$ (530 nm) and $T_{baseline}$ (530 nm) are the spectral transmittance of the soiled sample and clean coupon at 530 nm, respectively. The only difference between equation 2 and equation 6 is that transmittance is measured at a specific wavelength (530 nm).

Finally, electrical and transmittance losses due to soiling were calculated using equations 4, and equation 7, respectively:

$$T_{loss} \ (530 \ nm) \ [\%] = 100\% - T_{soiling} \ (530 \ nm). \tag{7}$$

## 4. Indoor thermal characterization

The purpose of the indoor thermal characterization of the components of DUSST was to quantify important temperature dependencies for use in correcting outdoor measurements to a nominal temperature. Well quantified temperature corrections serve to minimize measurement error as well as to provide opportunities for future designs that are more robust to varying outdoor temperatures.

Five tests were performed indoors under the previously mentioned identical conditions of temperature, humidity and light intensity to thermally characterize the different components of DUSST. The procedure followed for these tests is summarized in the flowchart of Fig. 7. The maximum standard deviation of the cell current measurements after the five tests was 2.3%.

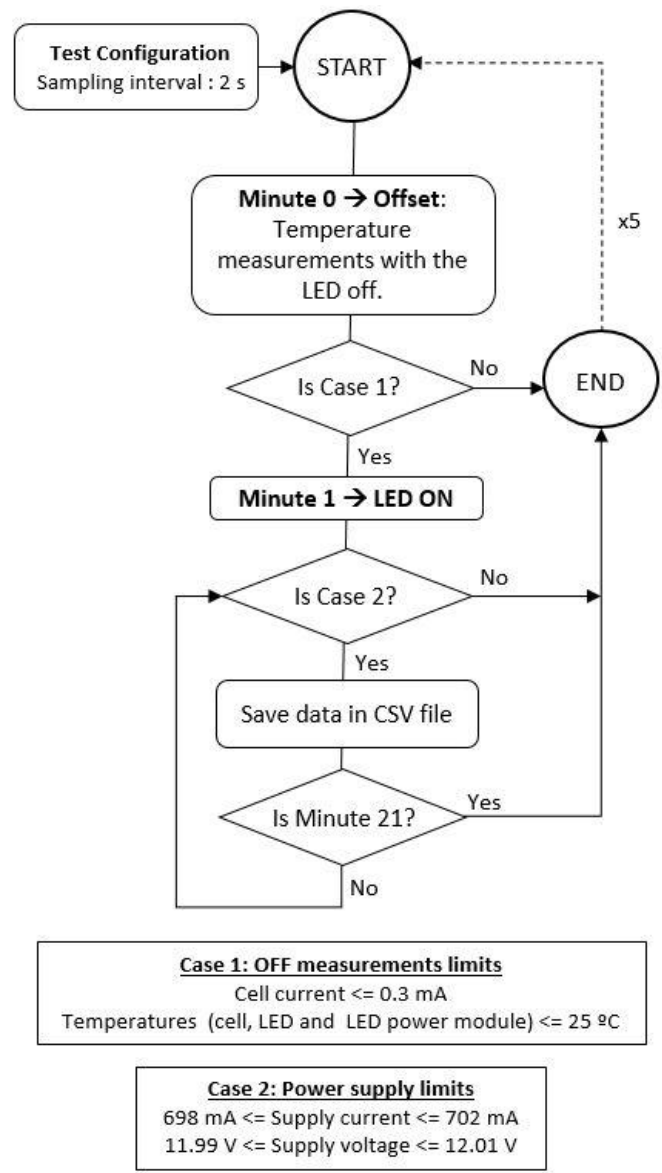

*Fig. 7. Flowchart of thermal characterization tests.*

The main results of this thermal characterization are presented and discussed in this section. Fig. 8 shows the evolution of the temperature of the different components with time, from the moment the measurement starts. As it can be seen, all of them follow a logarithmic trend, with the LED temperature experiencing the largest increase of 20 ºC in the first 10 minutes. In comparison, the LED power module (Buckpuck Temperature) increased ~10 ºC while the cell only increased 2 ºC. No relationships were found between the buckpuck temperature and LED current or voltage. Because of this, only the LED temperature is used to obtain a correlation with the electrical output of the PV cell. Future work will be conducted to analyse the impact of a wider temperature range on DUSST components, with a special focus on the buckpuck output.

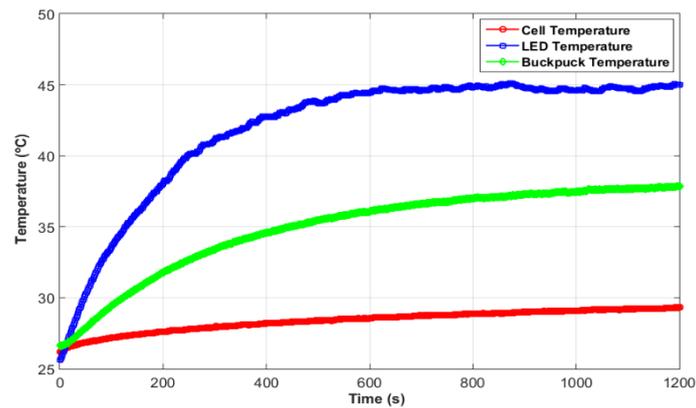

*Fig. 8. Evolution of components temperature with time.*

Fig. 9 shows the relationship found between LED voltage and temperature. The linear fit shows that LED voltage decreases with LED temperature with a linear rate of 2.54 mV/ºC, which is within manufacturer specifications.

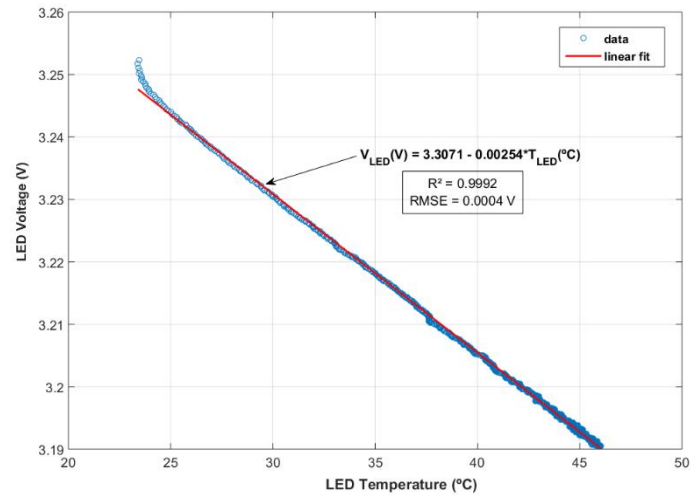

*Fig. 9. Linear correlation between LED Voltage and LED Temperature.*

The significant change in LED temperature and its effect on the LED voltage suggests that it will have an effect on the DUSST measurements, directly affecting the cell current. While the increase in cell temperature slightly increases the cell current output (short-circuit current increases with a rate of 0.06 %/ºC), its influence can be considered negligible in comparison with the effect of the rise in LED temperature. For this reason, the LED temperature was plotted against the cell current to find the correlation between them (Fig. 10). As shown in Fig. 10, the cell current decreases with increasing LED temperature at a linear rate of -0.052 mA/ºC. This relation has been named *TLED_Factor*. Because DUSST is designed to operate outdoors and to be exposed to a variety of climatic conditions, this finding will allow for correcting measurements at varying temperatures to a nominal operating temperature.

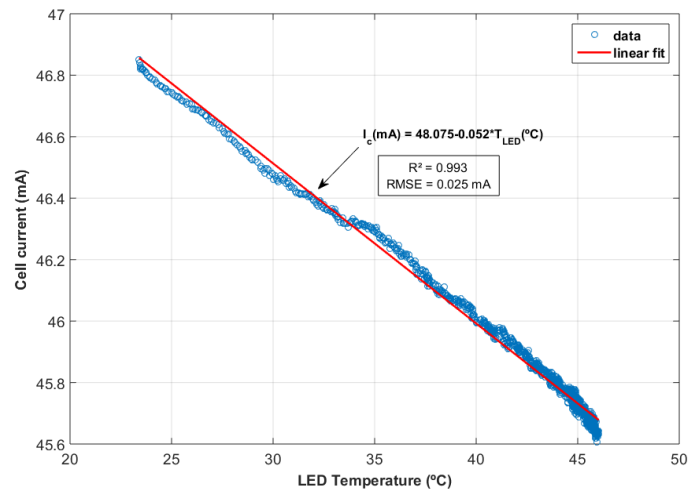

*Fig. 10. Linear correlation between PV cell current and LED temperature.*

## 5. Results and discussion

In this section, the results of an experimental campaign to obtain a model that can estimate accurately soiling transmittance losses are presented. Two distinct sets of transmittance and electrical measurements have been used first to build and validate the model, following the aforementioned methodology (Section 3.3) and thermal correction (Section 4). A discussion of the results of the two stages (calibration and validation) that compose this experimental analysis is also included.

Before the start of the calibration stage, the hemispherical transmittance (Fig. 11) of the reference coupon was measured at a wavelength of 530 nm ($T_{baseline}$ = 91.3%) and a DUSST measurement was performed to obtain the value of the PV cell current in clean conditions: $I_{baseline}$ = 43.4 mA.

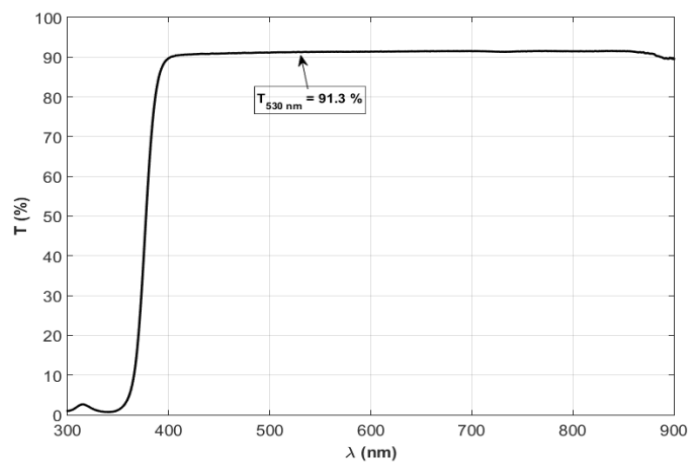

*Fig. 11. Baseline transmittance profile: PMMA PLEXIGLAS ® 8N clear coupon.*

Afterwards, measurements were taken for each of the 12 printed screening masks using both the spectrophotometer and DUSST. The recorded values were standardized using equations 3 and 6, and the transmittance and electrical losses were obtained through equations 4 and 7. The results can be seen in Table 2.

*Table 2. Calibration stage. Summary of transmittance and electrical losses.*

| Id | $T_{measured}$ (530 nm) | $T_{soiling}$ (530 nm) [Eq. 6] | $T_{loss}$ (530 nm) [Eq. 7] | I (mA) | LIR [Eq. 3] | DUSST$_{losses}$ [Eq. 4] |
|---|---|---|---|---|---|---|
| Mask 1 | 74.0% | 81.0% | **19.0%** | 38.89 | 89.8% | **10.2%** |
| Mask 2 | 64.2% | 70.3% | **29.7%** | 37.25 | 86.0% | **14.0%** |
| Mask 3 | 56.0% | 61.4% | **38.6%** | 35.79 | 82.6% | **17.4%** |
| Mask 4 | 48.6% | 53.3% | **46.7%** | 34.51 | 79.6% | **20.4%** |
| Mask 5 | 41.2% | 45.2% | **54.8%** | 33.22 | 76.7% | **23.3%** |
| Mask 6 | 34.3% | 37.6% | **62.4%** | 32.02 | 73.9% | **26.1%** |
| Mask 7 | 29.4% | 32.2% | **67.8%** | 30.31 | 69.9% | **30.1%** |
| Mask 8 | 28.0% | 30.7% | **69.3%** | 29.83 | 68.8% | **31.2%** |
| Mask 9 | 19.9% | 21.8% | **78.2%** | 23.74 | 54.8% | **45.2%** |
| Mask 10 | 16.8% | 18.4% | **81.6%** | 21.43 | 49.4% | **50.6%** |
| Mask 11 | 7.1% | 7.8% | **92.2%** | 11.56 | 26.7% | **73.3%** |
| Mask 12 | 3.2% | 3.5% | **96.5%** | 5.10 | 11.8% | **88.2%** |

Once transmittance and electrical losses were calculated, possible correlations were found by using a MATLAB® Toolbox (Curve Fitting Toolbox 3.5.1). Two feasible models were proposed: (1) a piecewise function consisting of two linear equations and (2) a logarithmic model, which are displayed in Fig. 12 and Fig. 13, respectively. To evaluate the accuracy of the models, two different statistical metrics have been used: the mean absolute error (MAE), also known as L1 loss, calculated through equation 8, and the root mean square error (RMSE), calculated by using equation 9.

$$MAE = \frac{\sum_{i=1}^{N}|x_{measured} - x_{modelled}|}{N}. \tag{8}$$

$$RMSE = \sqrt{\frac{1}{N}\sum_{i=1}^{N}(x_{measured} - x_{modelled})^2}, \tag{9}$$

where N is the number of data points, and $x_{measured}$ and $x_{modelled}$ represent the measured and the modelled data respectively. These metrics provide a way to measure the quality of the fit between the measured and the modelled values of losses by expressing the average model prediction error (the lower, the better). The main difference between them is that the RMSE is more sensitive to extreme outliers as the errors are squared before they are averaged. This can lead to a higher value of error, even if only a single modelled point differs from its respective measured point.

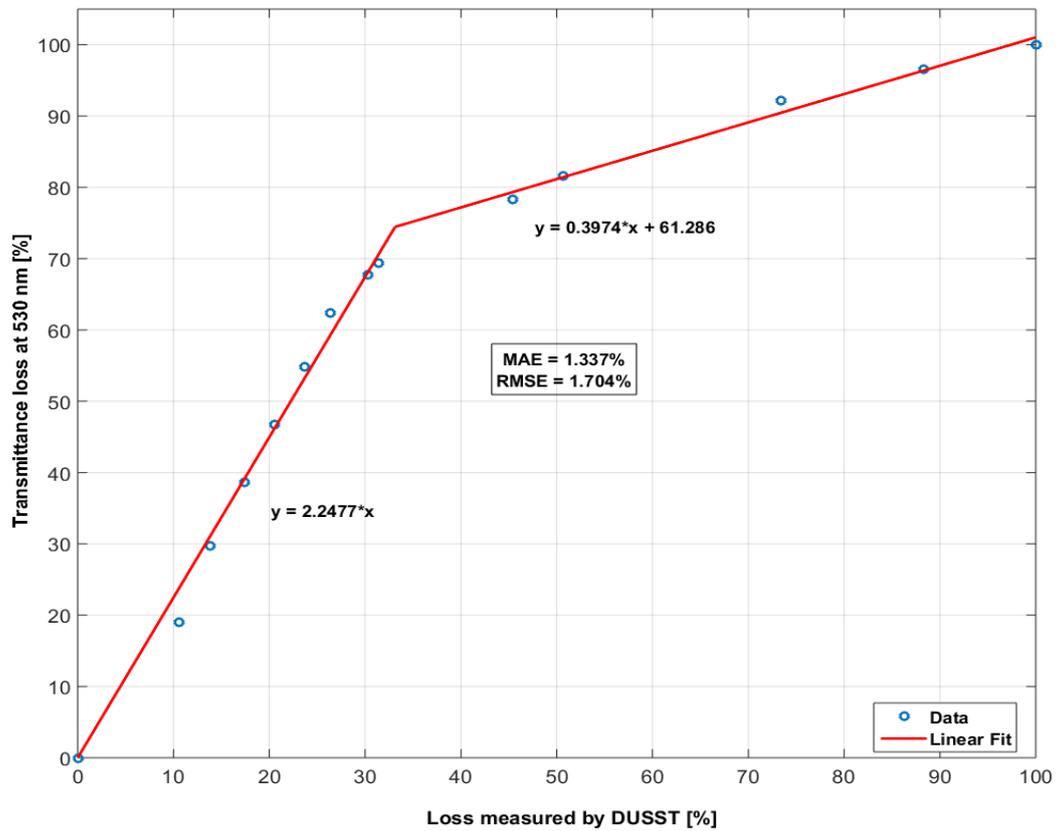

*Fig. 12. Correlation between transmittance and electrical losses. Piecewise linear model.*

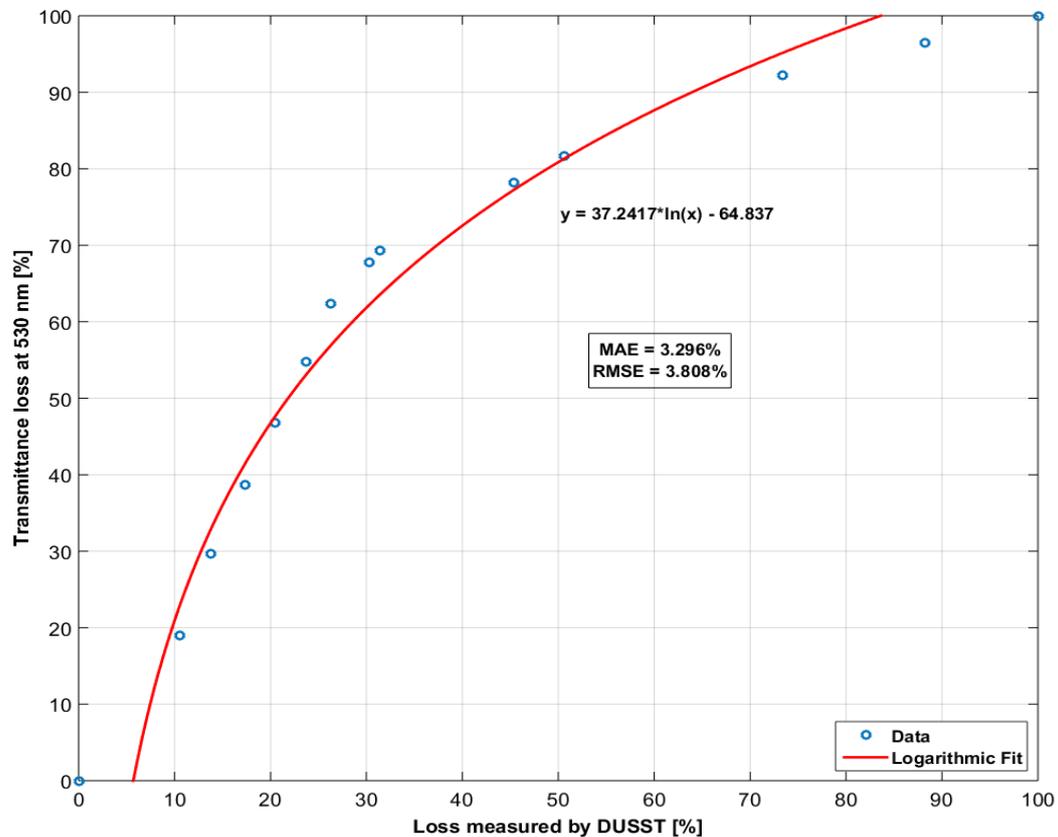

*Fig. 13. Correlation between transmittance and electrical losses. Logarithmic model.*

Table 3 presents the characteristics and the performance of the two models under investigation. Using two linear equations returns a RMSE of 1.7%, which is in line with the standard error found for other sensors (Gostein et al., 2019). Both models present values of MAE < 3.3% and of RMSE < 3.8%. Overall, the piecewise linear model returns the lowest RMSE (about a third of that found for the logarithmic fit): this is due to the fact that the logarithmic fit tends to overestimate the transmittance losses in the condition of intermediate soiling and underestimates the losses otherwise. For this reason, the logarithmic model was discarded in the following stage, as the magnitude of the errors increases when it is used to estimate soiling transmittance losses with naturally soiled coupons.

In the case of the piecewise linear model, it can be seen that the slope changes at a DUSST loss of ~33%. This could be justified by the low irradiance loss phenomenon, which affects the linear behavior of the electrical output of solar cells at the low irradiance conditions, values at which the DUSST cell is exposed (Hamadani et al., 2016; Mavromatakis et al., 2017). Future investigations should focus on the analysis of the solar detector to improve the linear response of DUSST for a wider range of intensity variations. In any case, as shown in Table 3, the piecewise linear model is found to quantify and correct this effect with a high accuracy.

*Table 3. Soiling transmittance models. Equations and statistical parameters.*

| Model | $T_{loss}(530\ nm)$ [%] | MAE | RMSE |
|---|---|---|---|
| Piecewise linear | $2.2477 \times DUSST_{losses}$; $\quad\quad\quad DUSST_{losses} \leq 33.1\%$ <br> $0.3974 \times DUSST_{losses} + 61.286$; $DUSST_{losses} > 33.1\%$ | 1.337% | 1.704% |
| Logarithmic | $37.2417 \times ln(DUSST_{losses}) - 64.837$ | 3.296% | 3.808% |

In the second stage of this analysis, the models just shown were validated against naturally deposited soiling. To do this, the outdoor soiled coupons were measured using the CaryWin spectrophotometer and DUSST. The above model was then used to convert DUSST measurements to transmittance losses for direct comparison with the direct measurements of the spectrophotometer. Due to the longer duration of this stage (from July 2019 to October 2019), baseline measurements for the clean coupon were conducted twice: at the beginning and at the end of the study. Only limited differences between measurements (< 0.5% for the transmittance and < 0.1 mA for the cell current) were found (see Table 4) and for this reason were expected not to influence the results.

*Table 4. Baseline measurements during the second stage. Validation.*

| Measurement | Start | End |
|---|---|---|
| $T_{baseline}$ | 91.3% | 91.0% |
| $I_{baseline}$ | 43.34 mA | 43.26 mA |

Transmittance measurements at 530 nm were taken for each soiled coupon, normalized through equation 6, and used to calculated the soiling transmittance losses by using equation 7 (See $T_{loss}$ in Table 5). The standard deviation (σ) reflects the level of non-uniformity of soiling on the coupon surface found after 4 measurements in different locations of each coupon.

*Table 5. Validation stage. Summary of measured transmittance losses with the spectrophotometer.*

| Id | $T_{soiling}$ (530 nm) [Eq. 6] | σ | $T_{loss}$ (530 nm) [Eq. 7] | Exposure time (days) |
|---|---|---|---|---|
| Coupon 1 | 93.2% | 0.6% | 6.8% | 13 |
| Coupon 2 | 85.7% | 3.3% | 14.3% | 20 |
| Coupon 3 | 86.4% | 5.0% | 13.6% | 25 |
| Coupon 4 | 78.6% | 9.9% | 21.4% | 53 |
| Coupon 5 | 85.9% | 0.1% | 14.1% | 63 |
| Coupon 6 | 86.3% | 0.9% | 13.7% | 73 |
| Coupon 7 | 83.1% | 3.0% | 16.9% | 77 |
| Coupon 8 | 80.3% | 7.6% | 19.7% | 90 |
| Coupon 9 | 79.0% | 3.8% | 21.0% | 97 |
| Coupon 10 | 84.2% | 0.9% | 15.8% | 103 |
| Coupon 11 | 88.0% | 2.5% | 12.0% | 111 |
| Coupon 12 | 86.8% | 1.2% | 13.2% | 118 |

Theoretically, if no precipitations had occurred during the experimental campaign, soiling transmittance losses would have increased with time. However, due to periodic rainfalls occurring between day 55 and the end of the campaign, the results did not follow this monotonic tendency. On the other hand, it should be highlighted that there are three coupons (3, 4, and 8) that showed a noticeable non-uniform accumulation of soiling, as the standard deviation of transmittance measurement exceed the value of 5%.

Following the procedure plotted in Fig. 5, the loss of each soiled coupon was measured with DUSST. As it was done with the data from the screening masks, equations 3 and 6 were applied to obtain the values of LIR and DUSST$_{losses}$ (Table 6). Finally, DUSST$_{losses}$ were translated to transmittance losses (T$_{loss, modelled}$) per the piecewise linear model in Table 3.

*Table 6. Validation stage. Summary of DUSST measurements.*

| Id | I (mA) | LIR [Eq. 3] | DUSST$_{losses}$ [Eq. 4] |
|---|---|---|---|
| Coupon 1 | 41.9 | 96.7% | 3.3% |
| Coupon 2 | 40.5 | 93.4% | 6.6% |
| Coupon 3 | 40.3 | 93.1% | 6.9% |
| Coupon 4 | 39.5 | 91.1% | 8.9% |
| Coupon 5 | 40.4 | 93.2% | 6.8% |
| Coupon 6 | 40.4 | 93.1% | 6.9% |
| Coupon 7 | 39.9 | 92.0% | 8.0% |
| Coupon 8 | 39.9 | 92.0% | 8.0% |
| Coupon 9 | 39.0 | 89.9% | 10.1% |
| Coupon 10 | 40.6 | 93.6% | 6.4% |
| Coupon 11 | 41.4 | 95.6% | 4.4% |
| Coupon 12 | 40.8 | 94.2% | 5.8% |

In Table 6, it can be seen that all the soiling deposited during the experimental campaign caused losses lower than 33.1%, so only the first equation of the piecewise linear model was applied. Fig. 14 shows the fit between the measured soiling transmittance losses and the estimated losses by using the piecewise linear model. The error bars represent the amount of non-uniform soiling over the coupons: the longer their length, the higher level of non-uniformity. As DUSST is not specifically designed to detect non-uniform soiling, these samples can influence the accuracy of the models. With an ideal model, the best fit would be a straight line with a slope of 1, and with values of coefficient of determination ($R^2$) and RMSE of 1 and 0, respectively.

Fig. 14 shows that the piecewise linear model presents a high accuracy with a slope of 0.987, a $R^2 = 0.943$ and a RMSE = 1.38%. As the slope is less than 1, this model slightly underestimates the losses. Nevertheless, it can be considered a suitable model to estimate soiling transmittance losses from DUSST measurements.

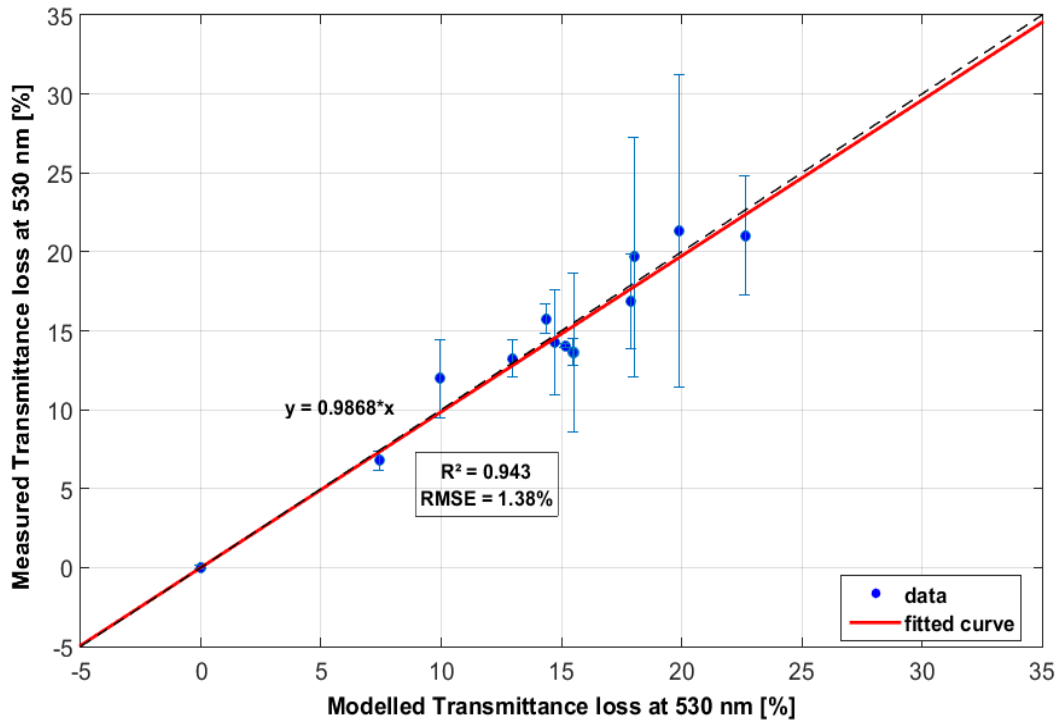

*Fig. 14. Piecewise linear model validation. Fit between measured (y-axis) and modelled (x-axis) transmittance losses.*

## 6. Conclusions

In this work, two analyses have been carried out to characterize the operation of the DUSST soiling sensor under controlled environmental conditions. Low cost soiling sensors such as DUSST offer the possibility to cost effectively estimate soiling losses at photovoltaic plants, and thus, to plan an optimal cleaning schedule in order to increase economic profits.

First, the thermal behaviour of the components of DUSST was studied, by analysing the different relationships between temperatures and electrical parameters. According to the results, the temperatures of the components increased logarithmically over time, with a stabilization time of 10 minutes after the LED was turned on. The temperature of the LED experienced the most notable increase and had the largest impact on the current generated by the PV cell. A linear relationship between these two parameters was found, where the cell current decreases with a factor of 0.052 mA/ºC with LED temperature. This correlation enables DUSST to be used under varying outdoor temperatures, as the measurement can be corrected according to the device's temperature. This will grant consistent LIR calculations independently of the outdoor temperatures. In future, an additional investigation should be conducted to evaluate how a wider range of temperatures will affect the electrical output of each DUSST component, including the buckpuck.

Subsequently, the possibility of estimating transmittance soiling losses using DUSST was investigated. The measurements of DUSST were compared with transmittance losses measured with a spectrophotometer. Screening masks were first used to develop a model to translate DUSST electrical measurements to transmittance losses; two feasible models (RMSE < 4%) were obtained: (1) a piecewise linear and (2) logarithmic. These models were then validated using naturally soiled coupons; the logarithmic model was discarded because its predictions fitted the data poorly. However, the piecewise linear model showed accurate results ($R^2$ > 0.94 and a slope almost equal to 1); this corroborates that it is viable to accurately estimate transmittance losses due to soiling through DUSST measurements. Notwithstanding this fact, the coupons were exposed to low-medium level of soiling, so the investigation was conducted only for losses up to 25%. Analogous investigations should be replicated in future at locations with higher soiling accumulations.

Future works will analyse and compare the soiling ratio values of different PV technologies through the measurements taken by DUSST and the model presented in this study. Hence, the reliability of both the detector and the model will be validated in a wider range of outdoor conditions. Additionally, the use of several prototypes based on light sources with LEDs emitting at different wavelengths will be investigated in order to model the complete spectral transmittance curve of soiling.

## Acknowledgements

Alvaro F. Solas, the corresponding author of this work, is supported by the Spanish ministry of Science, Innovation and Universities under the program "Ayudas para la formación de profesorado universitario (FPU), 2018 (Ref. FPU18/01460)". Eduardo F. Fernandez thanks the Spanish Ministry of Science, Innovation and Universities (RYC-2017-21910). Part of this work was funded through the European Union's Horizon 2020 research and innovation programme under the NoSoilPV project (Marie Skłodowska-Curie grant agreement No. 793120). This study is partially based upon work from COST Action PEARL PV (CA16235), supported by COST (European Cooperation in Science and Technology). COST (European Cooperation in Science and Technology) is a funding agency for research and innovation networks. Our Actions help connect research initiatives across Europe and enable scientists to grow their ideas by sharing them with their peers. This boosts their research, career and innovation, see www.cost.eu. Part of the work was authored by Alliance for Sustainable Energy, LLC, the manager and operator of the National Renewable Energy Laboratory for the U.S. Department of Energy (DOE) under Contract DEAC3608GO28308. Matthew Muller's work was supported by the DOE's Office of Energy Efficiency and Renewable Energy under Solar Energy Technologies Office Agreements 30311 and 34348.